\documentclass[aps,prl,showpacs,twocolumn]{revtex4-1}
\usepackage{amsmath,amssymb,latexsym,float}
\usepackage{fullpage,float,graphicx,mathrsfs}
\usepackage{hyperref}

\newcommand{\bpr}{\begin{proof}}
\newcommand{\epr}{\end{proof}}
\newcommand{\bsol}{\begin{proof}[Solution:]}
\newcommand{\ben}{\begin{enumerate}}
\newcommand{\een}{\end{enumerate}}
\newcommand{\bit}{\begin{itemize}}
\newcommand{\eit}{\end{itemize}}
\newcommand{\beq}{\begin{equation}}
\newcommand{\eeq}{\end{equation}}

\newcommand{\qqqquad}{\qquad\qquad} 

\begin{document}

\title{Discrete Breathers in a Nonlinear Polarizability Model of Ferroelectrics}
\author{C. Hoogeboom}
\author{P.G. Kevrekidis}
\affiliation{Department of Mathematics and Statistics, University of Massachusetts, Amherst MA 01003-4515, USA}
\author{A. Saxena}
\author{A. R. Bishop}
\affiliation{Center for Nonlinear Studies and Theoretical Division, Los
Alamos National Laboratory, Los Alamos, NM 87545 USA}
\date{\today}
\begin{abstract}
	We present a family of discrete breathers, which exists in a nonlinear polarizability model of ferroelectric materials. 
The core-shell model is set up in its non-dimensionalized Hamiltonian form and its linear spectrum is examined. Subsequently, seeking localized solutions in the gap of the linear spectrum, we establish that numerically exact and potentially stable discrete breathers exist for a wide range of frequencies therein.

	In addition, we present nonlinear normal mode, extended spatial profile solutions from which the breathers bifurcate, as well as other associated phenomena such as the formation of phantom breathers within the model. 
	The full bifurcation picture of the emergence and disappearance of the breathers is complemented by direct numerical simulations of their dynamical instability, when the latter arises.
\end{abstract}

\maketitle

\section{Introduction}

	The study of discrete breathers has gained considerable momentum over the past decade. 
	This has been motivated chiefly by their experimental observation in a wide variety of physical systems. 
	Such exponentially localized in space and periodic in time nonlinear waveforms have been identified in complex electronic materials such as halide-bridged transition metal complexes as, e.g., in \cite{swanson}; they have also 
been proposed to be relevant in the formation of denaturation bubbles in the DNA double strand dynamics as discussed, e.g., in \cite{peyrard}. 
	Such localization has also been discussed in the context of ultracold Bose-Einstein condensates immersed in optical lattices~\cite{oberthaler}. 
	Additional prototypical frameworks for the experimental formation of such modes include the nonlinear optics of evanescently coupled waveguide arrays, and biased photorefractive crystals; a recent review can be found in \cite{discreteopt}.  
	Localized modes have also been observed in micromechanical cantilever arrays~\cite{sievers}, in coupled torsion
pendula~\cite{lars}, in electrical lattices and transmission lines~\cite{lars2}, in granular crystals of beads interacting through Hertzian contacts~\cite{theo10}, and in layered antiferromagnetic samples such as those of a 
(C$_2$H$_5$NH$_3$)$_2$CuCl$_4$ crystal~\cite{lars3,lars4}.

	In the context of ferroelectrics, a prototypical model for their nonlinear dynamical description can be traced back to the work of Bilz et al. \cite{bilz87}. 
	It was subsequently shown that, independently of structural differences, this nonlinear polarizability model is applicable in general to displacive-type ferroelectrics including the temperature dependence of soft-mode frequencies 
and dielectric constants \cite{bussman-holder89}. 
	The theoretical analysis of this model has generally required considerable simplifications and assumptions \cite{bilz87}. 
	Yet, given the fundamental ingredients 
of nonlinearity and discreteness and their interplay relevant for
such systems,
discrete breathers (or intrinsic localized modes) have been discussed as 
potentially arising within this model.  
	They are associated with the formation of polar nanoregions 
indicating a crossover between and coexistence of displacive 
and relaxor behavior of ferroelectrics~\cite{bishop10,bussmann-holder04}.

	In this paper, our aim is to undertake a systematic numerical study of the existence and stability properties, as well as the nonlinear dynamics, of breathers within this polarization model for a 
prototypical set of parameter values, relevant to realistic contexts (ferroelectric materials). 
	We illustrate that breathers not only exist but can be linearly stable within the dispersion gap of the underlying linear system. 
	In the process, we also explore some of the interesting symmetry-breaking (pitchfork) bifurcations that give rise to the emergence of these waveforms within the nonlinear polarizability model and some of the (second harmonic generation) resonance phenomena that they incur.  
	Our presentation is structured as follows: In section II, we give the theoretical description of the model and 
its underlying linear spectrum. 
	In section III, we present our numerical results for the existence, stability and dynamical properties of the obtained discrete breathers. 
	Finally, in section IV, we summarize our findings and present some conclusions and suggestions for future studies.

\begin{figure}
\includegraphics[width=.40\textwidth]{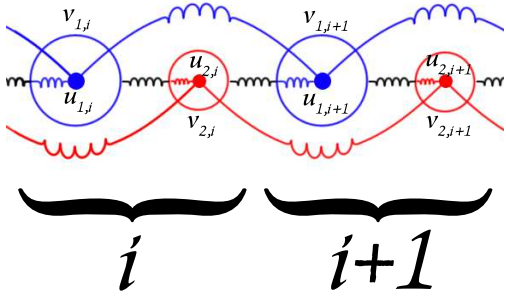}
\caption{
	The schematic of the pseudo-one dimensional nonlinear polarizability model for displacive-type ferroelectrics.  
	A diatomic lattice with core ($u_{1,j})$, $u_{2,j}$) and shell ($v_{1,j}$, $v_{2,j}$) displacements along with the connecting springs is shown for two neighboring sites $i$ and $i+1$. 
	} 
\label{f:schematic}
\end{figure}

\section{Theoretical Model}
	The nonlinear polarizability model describes a ferroelectric material by grouping atoms in the full system  into a ``pseudo-NaCl" structure, and then projecting it onto the direction of ferroelectric polarization 
\cite{bilz87}, simplifying a 3-D lattice to a diatomic 1-D lattice that retains many of the features of the original model. 
	Every lattice site has two atoms, each with a core and a shell (Fig. 1). 
	The displacements from equilibrium of cores one and two at lattice site $i$ are $u_{1,i}$ and $u_{2,i}$, and the displacements of the corresponding shells are $v_{1,i}$ and $v_{2,i}$. 
	The Hamiltonian of the model is given by
\begin{equation}
H=T+V, \label{e:hamiltonian}
\end{equation}
where
\begin{align}
T&=\frac{1}{2}\sum_i(m_1\dot{u}_{1,i}^2+m_2\dot{u}_{2,i}^2
+m_{e1}\dot{v}_{1,i}^2+m_{e2}\dot{v}_{2,i}^2),\label{e:kinetic}\\
V&=\frac{1}{2}\sum_i[f_1'(u_{1,i}-u_{1,i-1})^2+f_2'(u_{2,i}-u_{2,i-1})^2\nonumber\\
&\qqqquad +f(v_{2,i}-v_{1,i})^2+f(v_{2,i}-v_{1,i+1})^2\nonumber\\
&\qqqquad +g_2^{(1)}(v_{1,i}-u_{1,i})^2+g_2^{(2)}(v_{2,i}-u_{2,i})^2\nonumber\\
&\qqqquad +\tfrac{1}{2}g_4(v_{1,i}-u_{1,i})^4]. \label{e:potential}
\end{align}

\noindent Here $m_1$, $m_{e1}$,  $m_2$ and $m_{e2}$ denote the core and shell masses of atoms 1 and 2, respectively.  
	The shell-shell, core-core, and internal core-shell harmonic force constants are given by $f$, $f'_i$, and $g_2^{(i)}$ for $i=1,2$.  
	The anharmonic force constant $g_4$ takes into account the nonlinear polarizability of atom 1.  
	Typical values of the physical parameters, which were taken to match those of KTaO$_3$ in \cite{bussman-holder89} are shown in Table \ref{t:original_parameters}. 
\begin{table}
\begin{tabular}{l  r l  l r}
\hline
\hline
Parameter & Value & & Parameter & Value\\[3pt]
\hline\noalign{\smallskip}
$m_1$ (g)		&	$2.88\times 10^{-22}$	& &
						$m_2$ (g)		&	$0.649\times 10^{-22}$\\[3pt]
$m_{e1}$ (g)	&	$8.84\times 10^{-26}$	& &
						$m_{e2}$ (g)	&	$1.73\times 10^{-26}$\\[3pt]
$f'_1$ (gs$^{-2}$)			&	$0.581\times 10^4$	& &
						$f'_2$ (gs$^{-2}$)			&	$0.581\times 10^4$ \\[3pt]

$g_2^{(1)}$ (gs$^{-2}$)		&	$-0.828\times 10^4$	& &
						$g_2^{(2)}$ (gs$^{-2}$)		&	$1.00\times 10^4$\\[3pt]
$f$ (gs$^{-2}$)	&	$4.067\times 10^4$		& &
						$g_4$ (gs$^{-2}$cm$^{-2}$)	&	$0.255\times 10^{22}$\\[3pt]
$t_c$ (s)		&	$3.2675\times 10^{-15}$				& &
						$x_c$ (cm)					&	$.18020\times 10^{-8}$\\[3pt]
$e_c$ (gcm$^2$s$^{-2}$)		&	$2.6886\times 10^{-14}$&&&\\
\hline
\hline
\end{tabular}
\caption{
	Physical parameter values in Eqs. (2) and (3).
}\label{t:original_parameters} 
\end{table}

	It is convenient to rescale the Hamiltonian \eqref{e:hamiltonian} to the dimensionless form, so that the motion of shell-1 is of order unity.
	This can be achieved by making the substitutions $t=t_c\tau$, $u=x_c\mu$, and $v=x_c\gamma$, and $H=e_c\hat{H}$, where
\begin{align}
t_c=\sqrt{\frac{m_{e1}}{-g_2^{(1)}}}, \quad x_c=\sqrt{\frac{-g_2^{(1)}}{g_4}},
\quad e_c=\frac{(g_2^{(1)})^2}{g_4}.
\label{e:char_time_space}
\end{align}
	Now, for convenience, we drop the hats and relabel $\mu$ and $\gamma$ as $u$ and $v$; we will make sure, from this point forward, to specify when we are referring to the dimensional form. 
	Note here that any solution obtained in the dimensionless system can be re-dimensionalized by multiplying length by $x_c$, time by $t_c$, or energy by $e_c$; these are shown in Table \ref{t:original_parameters}.
	This gives us the resulting non-dimensional Hamiltonian
\begin{equation}
H=T+V,
\end{equation}
where
\begin{align}
T&=\frac{1}{2}\sum_i(\hat{m}_1\dot{u}_{1,i}^2+\hat{m}_2\dot{u}_{2,i}^2
+\dot{v}_{1,i}^2+\hat{m}_{e2}\dot{v}_{2,i}^2),\label{e:kinetichat}\\
V&=\frac{1}{2}\sum_i[C_1(u_{1,i}-u_{1,i-1})^2+C_2(u_{2,i}-u_{2,i-1})^2\nonumber\\
&\qqqquad +D(v_{2,i}-v_{1,i})^2+D(v_{2,i}-v_{1,i+1})^2\nonumber\\
&\qqqquad -(v_{1,i}-u_{1,i})^2+\alpha_2(v_{2,i}-u_{2,i})^2\nonumber\\
&\qqqquad +\tfrac{1}{2}(v_{1,i}-u_{1,i})^4],\label{e:potentialhat}
\end{align}
and
\begin{equation}
\begin{gathered}
C_1 = -\frac{f'_1}{g_2^{(1)}}, \qquad 
C_2 = -\frac{f'_2}{g_2^{(1)}},\\
D = -\frac{f}{g_2^{(1)}}, \qquad
\alpha_2 = -\frac{g_2^{(2)}}{g_2^{(1)}} \qquad, 
\hat{m}_1 = \frac{m_1}{m_{e1}}, \\
\hat{m}_2 = \frac{m_2}{m_{e1}}, \qquad
\hat{m}_{e2} = \frac{m_{e2}}{m_{e1}}.
\end{gathered}\label{e:dimensionless_parameters}
\end{equation}
This results in the following equations of motion
\beq
\begin{aligned}
\hat{m}_1\ddot{u}_{1,i}	  &= C_1(u_{1,i-1}+u_{1,i+1}-2u_{1,i}) \\
&\qquad								- (v_{1,i}-u_{1,i})
									+ (v_{1,i}-u_{1,i})^3 ,\\
\hat{m}_2\ddot{u}_{2,i}	  &= C_2(u_{2,i-1}+u_{2,i+1}-2u_{2,i})\\
&\qquad								+ \alpha_2(v_{2,i}-u_{2,i}) ,\\
\ddot{v}_{1,i}			  &= D(v_{2,n-1}+v_{2,n}-2v_{1,n})\\
&\qquad								+ (v_{1,n}-u_{1,n}) 
									- (v_{1,n}-u_{1,n})^3 ,\\
\hat{m}_{e2}\ddot{v}_{2,i}&= D(v_{1,n}+v_{1,n+1}-2v_{2,n})\\
&\qquad								- \alpha_2(v_{2,n}-u_{2,n}) .
\end{aligned}\label{e:dimensionless_motion}
\eeq
The schematic of this model is shown in Fig. \ref{f:schematic}. 
	Given the effective four-component nature of the model and the computational expense of performing the (up to a prescribed accuracy) exact numerical computation amounting to the identification of the breather solutions discussed
below, we have restricted our consideration to finite size lattices with $N$ sites and periodic boundary conditions ($u_{j,0}=u_{j,N+1}$, $v_{j,0}=v_{j,N+1}$, $j=1,2$). 
	Typically $N=20$ has been used in the results shown below, although we have ensured that this choice does not significantly affect the nature of our 
existence and stability results.

	Note also that this system is invariant under translations, so we have chosen (for ease of visualization) to shift each solution so that the spatial average of core-1 is zero. 
	Lastly, this system is clearly invariant under parity changes, so if the vector  $x=(u_{1,n},u_{2,n},v_{1,n},v_{2,n})$ represents a solution, then so does $-x$. 

\begin{figure}
\includegraphics[width=.45\textwidth]{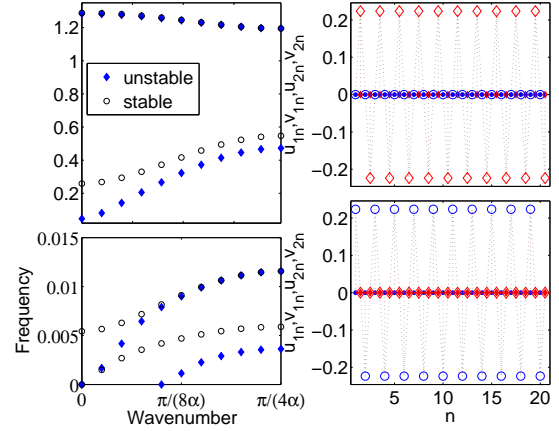}
\caption{
	The linear spectrum of both linearizations (i.e. stable and unstable equilibria) with four bands and three frequency gaps.  
	Note that the bottom two bands overlap.
	The right two panels show the eigenmodes from the bottom edge of the top band, and the top edge of the bottom band.
	In these panels, core/shell-1 are shown in blue circles, while core/shell-2 are shown as red diamonds. Cores are the filled markers, while shells are open markers.
	}\label{f:spectrum}
\end{figure} 

\subsection{Linear Spectrum}
	The above system possesses both stable and unstable uniform state equilibria. 
	Of the former kind are the solutions with  $u_{1,i}=0$, $u_{2,i}=v_{j,i}= \pm 1$, while of the latter
kind are the vanishing solutions with $u_{j,i}=v_{j,i}=0$.
	Equations \eqref{e:dimensionless_motion} were linearized about the equilibrium solutions to obtain the linearized equations $\ddot{x}=J(x^*)x$, where $J(x)$ is the Jacobian matrix of the system \eqref{e:dimensionless_motion}, and $x^*$
is the equilibrium vector being linearized around. 
	The linear spectrum changes depending on the particular form of $x$ about which we linearize.

	More specifically, we looked for solutions of the form $u_{1 n}=\xi_{1,n}\exp(i\omega t)$, $u_{2,n}=\xi_{2,n}\exp(i\omega t)$, $v_{1,n}=\nu_{1,n}\exp(i\omega t)$, and $v_{2,n}=\nu_{2,n}\exp(i\omega t)$, and then 
solved the ensuing eigenvalue problem.
	The linear spectrum of both the stable and unstable equilibria is shown in Fig. \ref{f:spectrum}.
	There are four bands and three frequency gaps (the bottom two bands overlap).
	The first band has frequencies in the range of 0 \textendash{} $5.894\times 10^{-3}$, and the dominant sublattice is that of core-1.
	The second band has frequencies in the range of $5.438\times 10^{-3}$ \textendash{} $11.57\times 10^{-3}$, and the dominant sublattice is that of core-2.
	The third band has frequencies in the range of 0.2589 \textendash{} 0.5473 Hz, and the dominant sublattice is that of shell-1.
	Lastly, the fourth band has frequencies in the range of 1.1949 \textendash{} 1.2885 Hz, and the dominant sublattice is that of shell-2.

\begin{figure*}
\includegraphics[width=.9\textwidth]{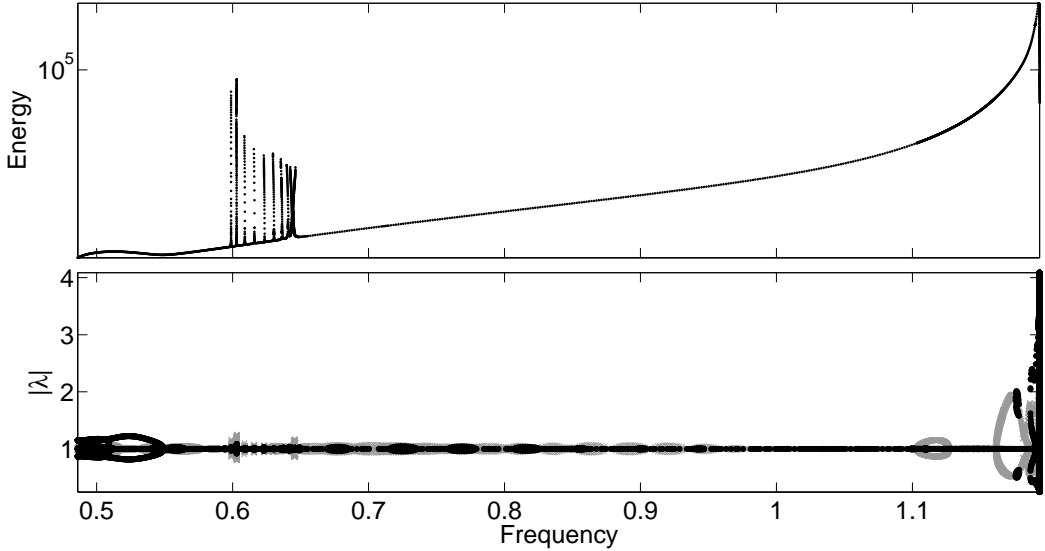}
\caption{
	\emph{Top:} the energy of the members of the breather solution branch is shown as a function of its frequency.
	Notice the {\it logarithmic} nature of the vertical scale.
	\emph{Bottom:} the Floquet multipliers of each breather solution are shown as a function of frequency. 
	Purely real Floquet multipliers deviating from the unit circle are shown in black, while complex multipliers with nonzero imaginary part deviating from the unit circle are shown in gray. 
	When neither of these deviations from unit modulus arises, the breather branch is linearly stable.
}\label{f:BB_energy_floq}
\end{figure*}

\section{Numerical Existence, Stability and Nonlinear Dynamics of Discrete Breathers}
We use a Newton-Raphson solver to identify the zeros of a Poincar\'{e} map 
for periodic orbits of the discrete breather type; for a relevant review
of computational methods for the identification of such states, see, e.g., 
\cite{flach20081}.
	Namely, we solve for the zeros of the map $P(x_0)=x_0-x(T,x_0)$, where $x(T,x_0)$ is a solution satisfying the equations of motion \eqref{e:dimensionless_motion} with initial condition $x_0$, evaluated at time $T$.
	Subsequently, a numerical continuation was performed using the breather frequency as the continuation parameter.
	At each continuation step we calculated the stability of the periodic solution in the form of its Floquet multipliers by computing the eigenvalues of the monodromy matrix $V(T)$, which is obtained by integrating the variational equation $V'=J(\bar{x}(t))V$ along the periodic solution $\bar{x}$ for one full period, with an initial condition of the identity matrix. 
	Floquet multipliers with a magnitude greater than one are associated with an unstable solution in our Hamiltonian system (where if $\lambda$ is a multiplier, then so are $1/\lambda$, $\lambda^{\star}$ and $1/\lambda^{\star}$).
	The eigenmode from the top edge of the unstable acoustic band was used as an initial guess, and the continuation was        performed for frequencies in the gap between the acoustic and optical bands.
	While this led to a family of time-periodic solutions, they were spatially extended and would not be classified as breathers (due to the absence of spatial localization).
	These had the same profile as the eigenmode from which they were spawned, and the amplitude increases as the frequency increases; such modes can be classified as nonlinear extensions of the (linear) normal modes of the system. 
Since this family bifurcates from the unstable equilibrium, we will call it the unstable branch (UB).
	
\begin{figure*}
\includegraphics[width=\textwidth]{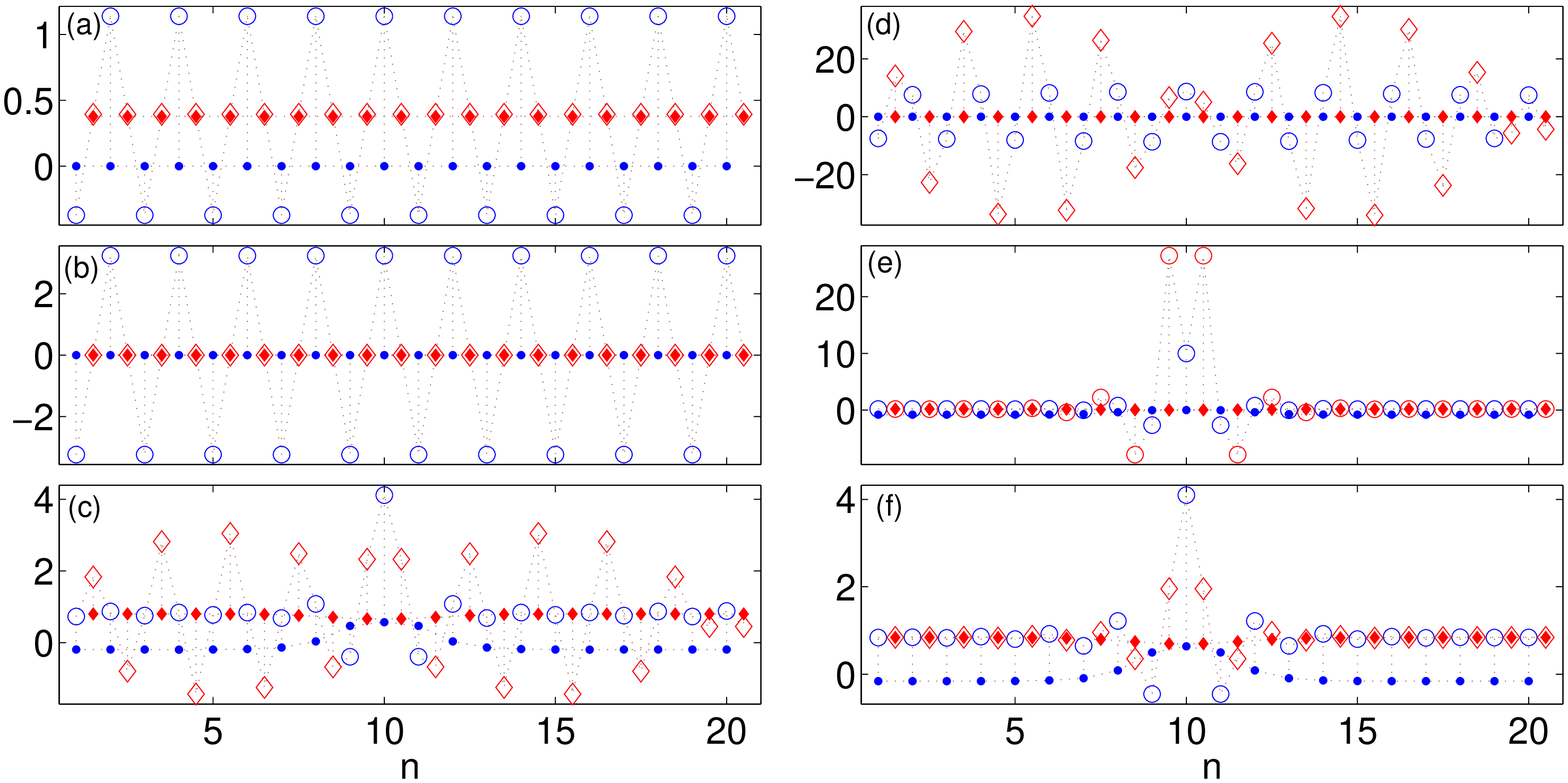}
\caption{
	Sample solutions from each continuation.
	\emph{(a)} An extended solution from the stable branch (SB).
	\emph{(b)} An extended solution from the unstable branch (UB).
	\emph{(c)} A nanopteron (phonobreather) from the breather branch (BB). 
	This particular solution has a frequency of 0.5989 and a second harmonic which is very close to the second frequency in the optical band.
	\emph{(d)} A breather solution from BB. This solution has a frequency (1.1936), which is very close to the bottom edge of the optical band. 
	Because of this, there are strong resonances with the corresponding optical mode.
	\emph{(e)} Another breather solution from BB. This solution is stable with a frequency of 1.1280.
	\emph{(f)} A third breather solution from BB. This solution has a frequency of 0.5884.
	In these panels, core/shell-1 are shown in blue circles, while core/shell-2 are shown as red diamonds. Cores are the filled markers, while shells are open markers.
}\label{f:sample_breathers}
\end{figure*}

Additionally, when a properly scaled eigenmode from the top edge of the stable equilibrium's acoustic band is added to the stable equilibrium, another family of spatially extended solutions can be identified. 
	This family originates at a frequency equal to the top of the stable acoustic band (0.5473), and continues down in frequency (up in energy) until it collides with UB at a subcritical (in the energy variable) pitchfork bifurcation.
	Since this family bifurcates from the stable equilibrium, we will call it the stable branch (SB). 
	Sample profiles of the stable and unstable branches (SB and UB, respectively), as well as the breather branch to which we turn to below, are displayed in Fig.~\ref{f:sample_breathers}.
	
	 In order to identify localized solutions, we needed a suitable initial guess.
	 We excited the shell of the first mass within a single lattice site in a chain with a total of 200 sites, and then integrated it in the equations of motion for a sufficient length of time that the extended excitations (phonons) were allowed to radiate away, resulting in a localized oscillation centered at the excited node. 
	 This provided us with a sufficiently good initial guess for the profile (truncated to only $N=20$ sites) and a reasonable approximation to the frequency of an exact breather solution.
	 This successfully converged to a breather with the Newton-Raphson method, which allowed a continuation through the gap.
	 There are many interesting features to this family of breathers, which we call the breather branch (BB) (see again
members of this family in  Fig.~\ref{f:sample_breathers}).
	 Figure \ref{f:BB_energy_floq} shows a view of the energy of each solution within this BB as a function of frequency and illustrates their corresponding stability. 
	 It is noticeable that, with the exception of a strong instability in the vicinity of the optical band and a weaker one in the vicinity of the acoustic band, this branch is generally found to be robust, and for a wide parametric regime of frequencies it is linearly stable.

\begin{figure}
\includegraphics[width=.5\textwidth]{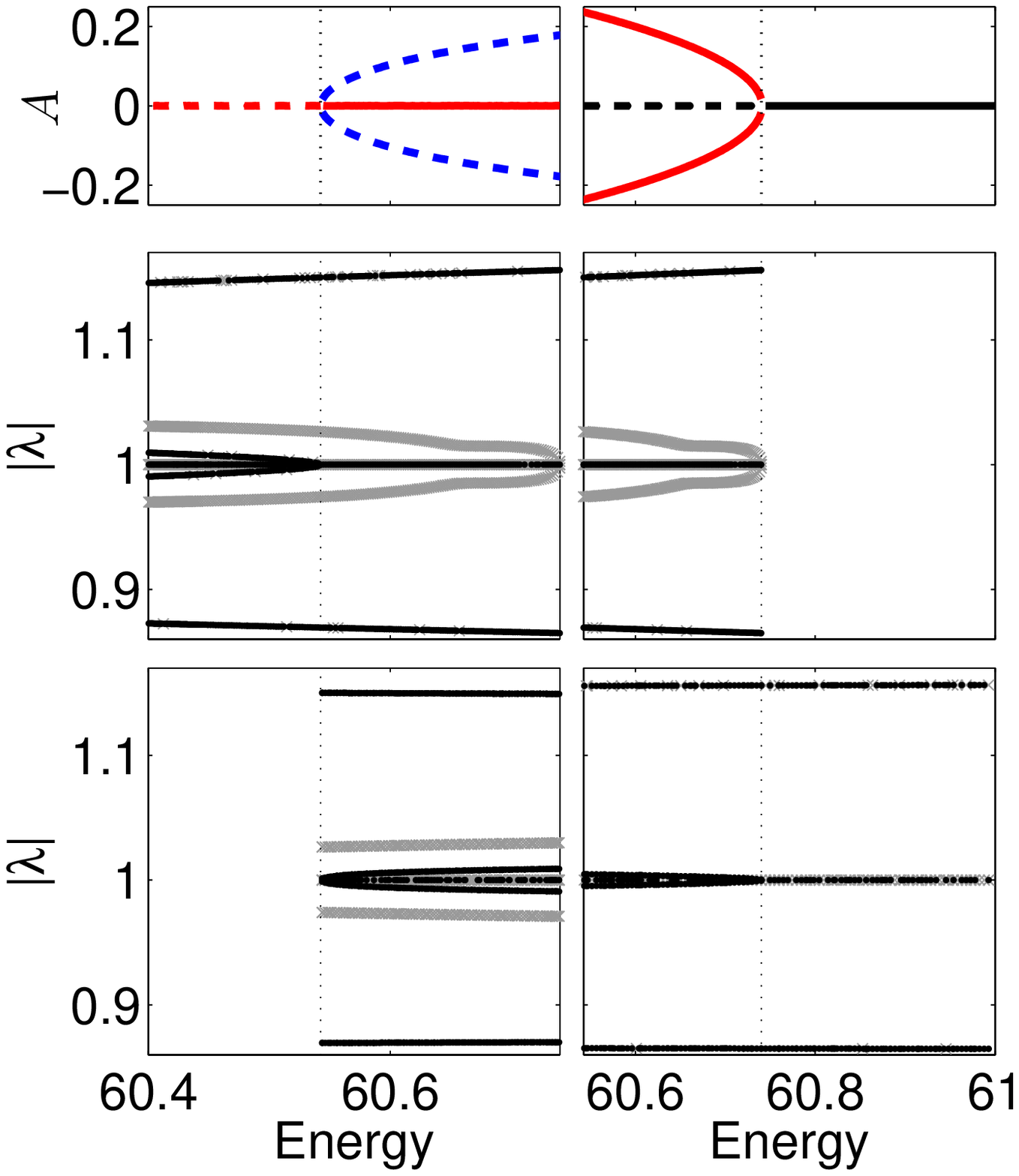}
\caption{
	\emph{Left:} the supercritical (in the energy variable) pitchfork bifurcation between SB and BB. 
	The top panel shows the relevant `symmetry' parameter as a function of energy. 
	The middle panel shows the Floquet multipliers of SB as a function of energy, and the bottom panel shows the Floquet multipliers of BB as a function of energy.
	\emph{Right:} the subcritical (again in the energy variable) pitchfork bifurcation between UB and SB.
	The top panel shows the `symmetry' parameter as a function of energy.
	The middle panel shows the Floquet multipliers of SB as a function of energy, and the bottom panel shows the Floquet multipliers of UB as a function of energy.
}\label{f:lower_bifurcation}
\end{figure}

\begin{figure}
\includegraphics[width=.45\textwidth]{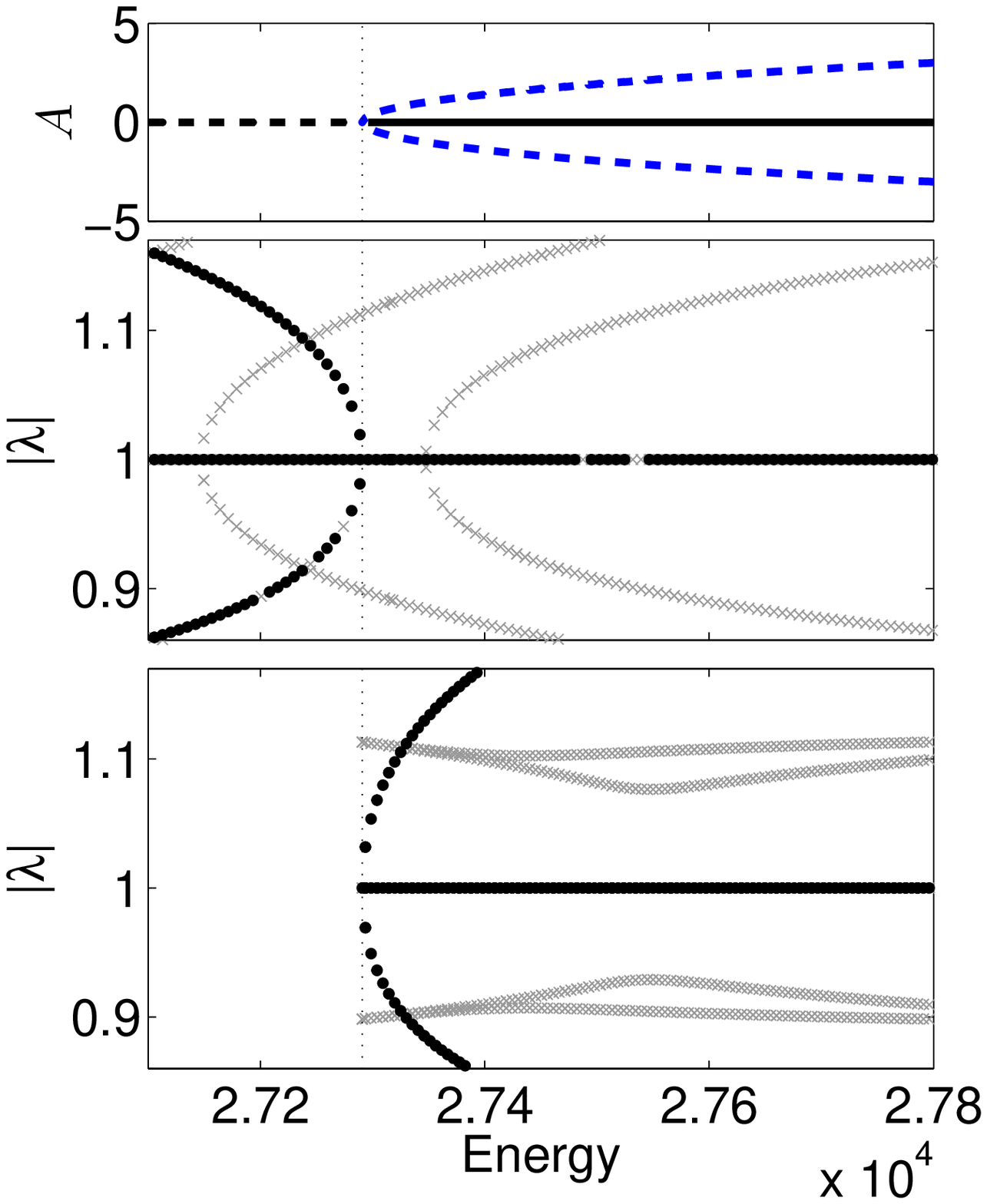}
\caption{
	The subcritical pitchfork bifurcation between BB and UB (for high energies and frequencies near the termination point of the breather branch).
	\emph{Top:} the `symmetry' parameter as a function of energy.
	\emph{Middle:} the Floquet multipliers of UB as a function of energy.
	\emph{Bottom:} the Floquet multipliers of BB as a function of energy.
}\label{f:upper_bifurcation}
\end{figure}

The figures~\ref{f:lower_bifurcation} and \ref{f:upper_bifurcation} illustrate the bifurcation events associated (directly or indirectly) with the breather branch.
	 BB emerges out of a supercritical (in the energy variable) pitchfork bifurcation of the SB, as is indicated in the left panel of Fig.~\ref{f:lower_bifurcation} [and thus inherits the (in)stability of this branch]. 
	 On the other end, for high frequencies, a turning point emerges (where the change of monotonicity is associated with a stability change) near frequency $\omega=1.2$.
	Further increase of frequency then leads to a rapid decrease in energy, and finally the branch merges with the UB in a pitchfork bifurcation that is subcritical in the frequency variable but supercritical in the energy variable.
	Notice that the Fig.~\ref{f:lower_bifurcation} in its right panel also illustrates the subcritical (in the energy variable) collision event between the UB and SB branches.

	 Notice also that in order to visualize each of the above mentioned bifurcations and to reveal their pitchfork (symmetry-breaking) character, we needed to select a proper measure of ``asymmetry'' $A$. 
	 In each case, $A$ was chosen based on how the profiles changed before and after the bifurcation. 
	 For instance, let us consider the bifurcation between SB and BB.
	 The defining difference between SB and BB close to the bifurcation is that solutions in SB have a core-1 that is constant across lattice sites, while solutions in BB have a core-1 that displays the (associated with the breather nature) spatial localization.
	 Here, we chose to have $A$ measure the sample skewness of the first core in our system of two cores-shells, as follows:
\beq
A=\frac{\tfrac{1}{N} \sum_{i=1}^N (u_{1,i}-\overline{u}_1)^3}{\left(\tfrac{1}{N} \sum_{i=1}^N (u_{1,i}-\overline{u}_1)^2\right)^{3/2}},
\eeq
where $\overline{u}_1$ is the mean value of core-1.
	The top-left panel of Fig.~\ref{f:lower_bifurcation} shows a plot of $A$ vs. energy, which explicitly illustrates the supercritical pitchfork structure of the bifurcation.
	In addition we see an exchange in stability of each branch that is characteristic of the pitchfork bifurcation.
	Although both branches are unstable before and after the bifurcation (due to a Floquet multiplier with modulus greater than one), we see that SB loses a real pair of Floquet multipliers, while the BB gains one.
	
\begin{figure*}
\includegraphics[width=.9\textwidth]{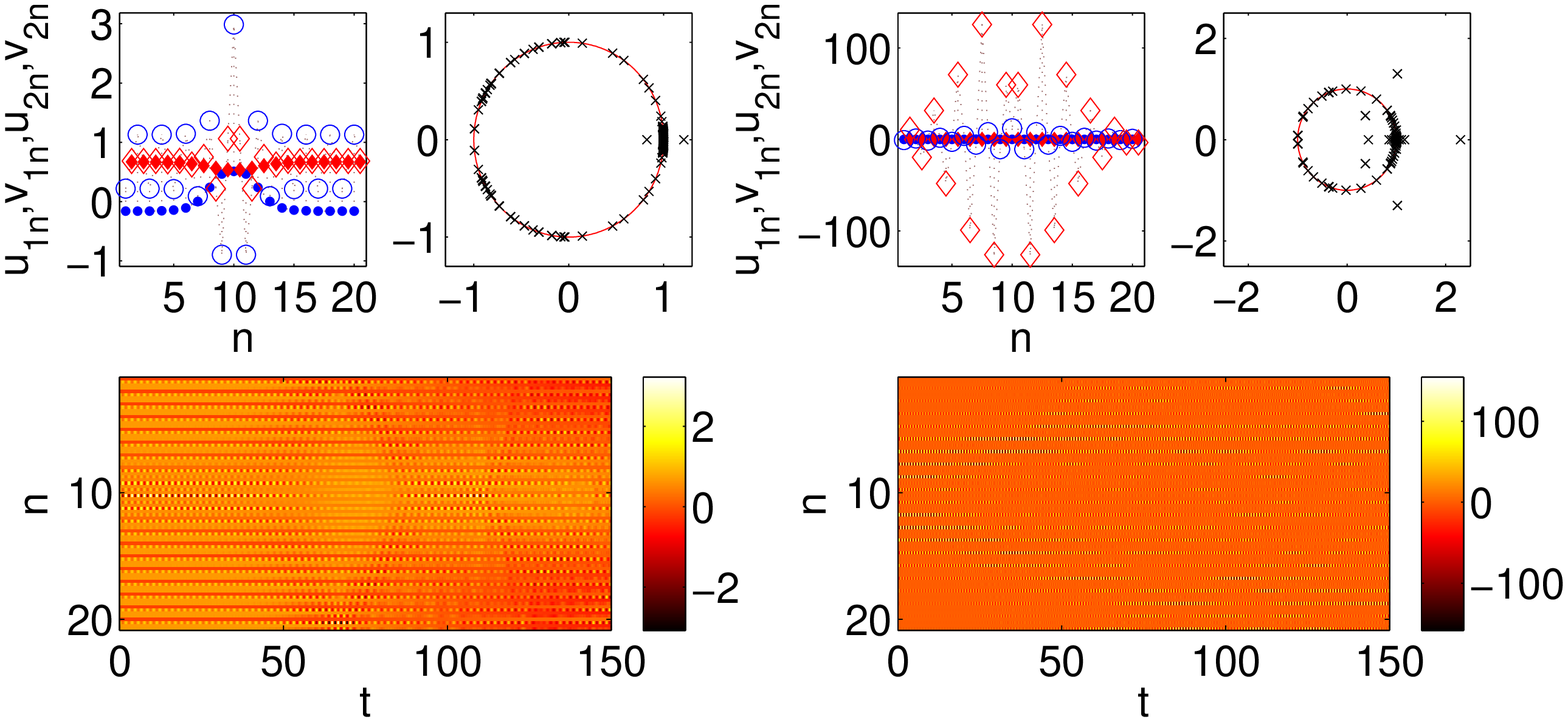}
\caption{
	\emph{Left:} the dynamics of an unstable breather with a frequency of 0.5292. 
	\emph{Right:} the dynamics of an unstable breather with a frequency of 1.1903.
	In these panels, core/shell-1 are shown in blue circles, while core/shell-2 are shown as red diamonds. Cores are the filled markers, while shells are open markers.
	In each case, the top two panels show the profile of the breather on the left and the Floquet multipliers on the right, while the bottom panel shows a spacetime plot of the displacements of this breather with a small perturbation.
}\label{f:unstable_dynamics}
\end{figure*}	
	
	Next, consider the bifurcation between SB and UB.
	The defining difference between SB and UB close to the bifurcation is that solutions in UB have homogeneous cores, while solutions in SB have both cores (one and two) which are displaced from each other. 
	In this case, we choose $A$ to be the mean of all four sublattices.
	The top-right panel of Fig. \ref{f:lower_bifurcation} shows a plot of $A$ vs. energy, which illustrates the subcritical pitchfork bifurcation.
	Once again, although both branches are unstable near the bifurcation, we see an exchange of a pair of Floquet multipliers that is characteristic of a pitchfork bifurcation; UB loses an unstable Floquet multiplier inheriting the 
stability (in this particular subspace) of the SB.
	
	Lastly, consider the bifurcation between BB and UB. Here, the defining difference between the two branches close to the bifurcation is that solutions in UB have a  homogeneous shell-2 in its spatial distribution, while the spatial localization associated with the BB branch leads to an inhomogeneous distribution of this field in the latter case.
	Hence, we have chosen $A$ to be the value of $v_2$ at the site with maximum amplitude.
	The top panel of Fig. \ref{f:upper_bifurcation} shows a plot of $A$ vs. energy, which illustrates the supercritical (in the energy variable) pitchfork bifurcation.
	Lastly, we see in the lower panels of Fig. \ref{f:upper_bifurcation} that BB inherits an unstable Floquet multiplier from UB.
	
	As one continues along BB from the lower bifurcation, the energy increases slightly, and then decreases to a local minimum, which occurs very near the frequency corresponding to the upper edge of the stable acoustic band.
	 Further along BB, we can observe in Fig.~\ref{f:BB_energy_floq} a series of spikes in the energy of the BB branch. 
	 These spikes correspond to frequencies for which the second harmonic of the breather is a frequency in the discrete linear band (and are thus associated with the corresponding resonances).
	 The type of solution that exists around these spikes has been termed a \emph{phonobreather}~\cite{morgante02}   (also occasionally referred to as \emph{nanopteron} due to the resonantly excited tail with non-vanishing amplitude).
	 These solutions comprise a localized breather solution oscillating at the fundamental frequency, superposed with a phonon solution oscillating at the second harmonic frequency. 

	 A sample profile of one such solution is shown in Fig. \ref{f:sample_breathers}.
	 Since the linear spectrum is so sparse for a system of only 20 lattice sites, we can continue past these points of resonance by using a step size large enough to skip over the exact resonance points, and to find breathers for  frequencies between them.
	 These breathers, termed \emph{phantom breathers}, exist in frequency intervals that should be forbidden for localized solutions in an infinite chain (where the linear spectrum becomes dense), as discussed more thoroughly in \cite{morgante02}. 
	 Once past the second harmonic resonances, there is a long range of frequencies for which there exist stable, strongly localized (breather) solutions.
	 As the frequencies approach the bottom edge of the optical band, the energy of the solution increases rapidly until it eventually reaches the turning point discussed above (and changes its stability, as expected by the energy-frequency monotonicity criterion; see also~\cite{theo10}) and the BB eventually merges with UB as illustrated above.	 
	 
	Having explored the existence and stability properties of the discrete breather solutions emerging in the nonlinear polarizability model of cores and shells in a ferroelectric material, our intention is to complement these findings by an examination of the nonlinear dynamics of such solutions, in the regimes where they are found to be unstable. 
	This is particularly relevant to explore in the two regimes where strongly unstable dynamics emerge, namely for low frequencies and energies, in the vicinity of the acoustic band, as well as for high frequencies and energies near the optical band.  
	 This is summarized in Fig. \ref{f:unstable_dynamics}.
	Interestingly, it is found that the breather dynamics is quite similar in its nature in both of these highly unstable regimes. 
	In particular, we observe that both breathers undergo about 25 full oscillations before eventually dispersing into linear waves, i.e., the BB solutions are completely destroyed by the dynamical instability. 
	Nevertheless, we emphasize once again that in a large fraction of the linear gap (and away from the edges of their existence region) the BB solutions are found to be very robust in the long-time numerical simulations that we have performed.

\section{Conclusion and Further Work}

	In the present work, we have been able to provide a direct numerical verification of the existence of discrete breathers in the nonlinear model for polarizability that has been proposed for displacive-type ferroelectric materials \cite{bilz87}. 
	We have identified the linear spectrum of the underlying model and have shown how breather type excitations can emerge within the gaps of the linear spectrum. 
	Additionally, we have provided a detailed continuation of such solutions within the gap that revealed a number of interesting characteristics. 
	In particular, we observed the existence of pitchfork-type symmetry breaking bifurcations near the edges of the acoustic and optical bands of the spectrum, that led to the emergence/termination of the breathers, as well as the formation of phantom breather solutions \cite{morgante02} in the vicinity of the second harmonic resonances of the breather frequency with the linear optical band modes. 
	It is also noteworthy that although the breathers appear to be destroyed by strong instabilities associated with real Floquet multipliers in the vicinity of the band edges of the gap, they in fact appear to be rather stable within a wide parametric (frequency) interval in the gap. 

	There are many interesting directions along which the present study can be extended. 
	On the one hand, it would be relevant to understand if the above identified breather states (and nonlinear normal modes of the unstable or of the stable type) are the only nonlinear waveforms that the 1-D system supports. 
	On the other hand, it is relevant to consider multi-dimensional generalizations of the model which may bear interesting features in their own right, such as the potential for discrete vortices and other genuinely higher-dimensional excitations. 
	Such avenues will be explored in future publications.

	This study provides solutions and properties of discrete breathers that can be tested experimentally for ferroelectric materials exhibiting coexistence of displacive behavior and polar nanoregions (e.g., for relaxor behavior) and probably a crossover between the two regimes~\cite{bishop10,bussmann-holder04}. 
	In particular, the presence of discrete breathers should be reflected in the temperature and frequency dependence of vibrational and dielectric properties~\cite{bishop10,bussmann-holder04}.

\bibliographystyle{aipnum4-1}
\bibliography{ferroelectrics}

%
%
%
%
%
%
%
%
%
%
%
%
%
\end{document}